\documentclass[twocolumn,american,english]{revtex4}
\usepackage[T1]{fontenc}
\usepackage[latin9]{inputenc}
\usepackage{color}
\usepackage{array}
\usepackage{units}
\usepackage{amsmath}
\usepackage{graphicx}
\usepackage{esint}

\makeatletter

\providecommand{\tabularnewline}{\\}

\@ifundefined{textcolor}{}
{%
 \definecolor{BLACK}{gray}{0}
 \definecolor{WHITE}{gray}{1}
 \definecolor{RED}{rgb}{1,0,0}
 \definecolor{GREEN}{rgb}{0,1,0}
 \definecolor{BLUE}{rgb}{0,0,1}
 \definecolor{CYAN}{cmyk}{1,0,0,0}
 \definecolor{MAGENTA}{cmyk}{0,1,0,0}
 \definecolor{YELLOW}{cmyk}{0,0,1,0}
}

\makeatother

\usepackage{babel}
\begin{document}

\title{\textcolor{black}{Molecular electric moments calculated by using
natural orbital functional theory}}

\author{\textcolor{black}{Ion Mitxelena$^{1,2}$, Mario Piris$^{1,2,3}$ }}

\address{\textcolor{black}{$^{1}$Kimika Fakultatea, Euskal Herriko Unibertsitatea
(UPV/EHU), P.K. 1072, 20080 Donostia, Spain.}}

\address{\textcolor{black}{$^{2}$Donostia International Physics Center (DIPC),
20018 Donostia, Spain.}}

\address{\textcolor{black}{$^{3}$IKERBASQUE, Basque Foundation for Science,
48013 Bilbao, Spain.}}
\begin{abstract}
\textcolor{black}{The molecular electric dipole, quadrupole and octupole
moments of a selected set of 21 spin-compensated molecules are determined
employing the extended version of the Piris natural orbital functional
6 (PNOF6), using the triple-$\zeta$ Gaussian basis set with polarization
functions developed by Sadlej, at the experimental geometries. The
performance of the PNOF6 is established by carrying out a statistical
analysis of the mean absolute errors with respect to the experiment.
The calculated PNOF6 electric moments agree satisfactorily with the
corresponding experimental data, and are in good agreement with the
values obtained by accurate }\textit{\textcolor{black}{ab initio}}\textcolor{black}{{}
methods, namely, the coupled-cluster single and doubles (CCSD) and
multi-reference single and double excitation configuration interaction
(MRSD-CI) methods.}
\end{abstract}
\maketitle

\section{\textcolor{black}{Introduction}}

\textcolor{black}{The interpretation and understanding of intermolecular
forces, particularly those relating to long-range electrostatic interactions,
require knowledge of the electrostatic moments \cite{Buckingham1959,Buckingham1967}.
The electric moments are essential to provide simple ways to figure
out the electric field behaviour of complex molecules. These electrical
properties provide also information about the molecular symmetry since
the electric moments depend on the geometry and charge distribution
of the molecule.}

\textcolor{black}{It has long recognized the role of electrostatic
interactions in a wide range of biological phenomena \cite{Honig1995}.
The electrostatic energy is frequently the ruling contribution to
molecular interactions in large biological systems, hence it is extremely
important to describe properly the electrostatic potentials around
these molecules. In order to improve the current treatment of the
electrostatics for biomolecular simulations, which are traditionally
modeled using a set of atom-centered point charges, the knowledge
of higher multipole moments is required to include the effects of
non-spherical charge distributions on intermolecular electrostatic
interactions.}

\textcolor{black}{In principle, one can experimentally find the components
of the electric field at each point, but it turns into a formidable
task for large molecular systems. There are several techniques to
determine experimentally the dipole moments \cite{GordyEXP,McClellanEXP},
but it is still very difficult to obtain precise experimental values
of higher multipole moments such as quadrupole or octupole moments
\cite{Buckingham1959,FlygareEXP,CohenEXP}, independently of the experimental
conditions. Theoretical calculations are therefore essential but challenging
for quantum chemistry methods. The accurate calculation of these properties
is highly dependent on the method employed \cite{Glaser}, either
regarding approximate density functionals \cite{CohenDFT} or methods
based on wavefunctions \cite{Bundgen,Junquera}. Consequently, calculating
the multipole moments is a way to assess any electronic structure
method.}

\textcolor{black}{The natural orbital functional (NOF) theory \cite{Piris2007}
has emerged in recent years \cite{Piris2014a,Pernal2016} as an alternative
method to conventional }\textit{\textcolor{black}{ab initio}}\textcolor{black}{{}
approaches and density functional theory (DFT). A series of functionals
has been proposed by Piris and collaborators (PNOFi, i=$\overline{1,6}$)
\cite{Piris2013b,Piris2014} using a reconstruction of the two-particle
reduced-density matrix (2-RDM) in terms of the one-particle RDM (1-RDM)
by ensuring necessary N-representability positivity conditions on
the 2-RDM \cite{Piris2006}. In this work, we employ the PNOF6 \cite{Piris2014},
which has proved a better treatment of both dynamic and non-dynamic
electron correlations than its predecessors \cite{Ramos-Cordoba2015,Piris2015a,Lopez2015,Cioslowski2015c,Piris2016}.}

\textcolor{black}{The aim of the present paper is to apply PNOF6,
in its extended version, to the determination of molecular dipole
and quadrupole moments of selected spin-compensated molecules, namely,
H$_{2}$, HF, BH, HCl, H$_{2}$O, H$_{2}$CO, C$_{2}$H$_{2}$, C$_{2}$H$_{4}$,
C$_{2}$H$_{6}$, C$_{6}$H$_{6}$, CH$_{3}$CCH, CH$_{3}$F, HCCF,
ClF, CO, CO$_{2}$, O$_{3}$, N$_{2}$, NH$_{3}$, and PH$_{3}$.
Moreover, the octupole moment of CH$_{4}$, a molecule without dipole
and quadrupole moments is also studied. The Gaussian basis set of
Sadlej \cite{Sadlej1988,Sadlej1991}, which has been specially developed
to compute accurately molecular electric properties, is employed to
perform all calculations.}

\textcolor{black}{We compare the obtained PNOF6 results with the experimental
values reported in the literature \cite{Bundgen,Junquera,Fabricant_ClF,FlygareEXP,Maroulis_ozone,Russel_CH3F,heard_benzene},
as well as with the theoretically computed values of Bundgen et al.
who used the multi-reference single and double excitation configuration
interaction (MRSD-CI) method, and the coupled-cluster single and doubles
(CCSD) values calculated by us. Recall that the CCSD values for one-electron
properties differ from full-CI results in only $2\%$ if no multiconfigurational
character is observed \cite{HalkierBH_HF}, so they can be considered
as benchmark calculations. To our knowledge, this is the first NOF
study of higher multipole moments such as quadrupole and octupole
moments.}

\textcolor{black}{This article is organized as follows. We start in
section II with the basic concepts and notations related to PNOF6
and electric multipole moments. The section III is dedicated to present
our results and those obtained by using CCSD and MRSD-CI methods.
Here, we discuss the outcomes obtained for the dipole, quadrupole
and octupole principal moments, in separate sections. The performance
of PNOF6 is established by carrying out a statistical analysis of
the mean absolute errors (MAE) with respect to the experimental marks.}

\section{\textcolor{black}{Theory}}

\subsection{\textcolor{black}{The NOF Theory}}

\textcolor{black}{We briefly describe here the theoretical framework
of our approach. A more detailed description of PNOF6 can be found
in reference \cite{Piris2014}. We focus on the extended version of
PNOF \cite{Piris2013e}, which provides a more flexible description
of the electron pairs in the NOF framework.}

\textcolor{black}{Recall that PNOF6 is an orbital-pairing approach,
which is reflected in the sum rule for the occupation numbers, namely,
\begin{equation}
\sum_{p\in\Omega_{g}}n_{p}=1\:,\quad g=\overline{1,F}\label{sumrule_n}
\end{equation}
}

\textcolor{black}{where $p$ denotes a spatial natural orbital and
$n_{p}$ its occupation number. This involves coupling each orbital
$g$, below the Fermi level ($F$), with $N_{c}$ orbitals above it
$\left(p>F\right)$, so the orbital subspace $\Omega_{g}\equiv\left\{ g,p_{1},p_{2},\cdots,p_{N_{c}}\right\} .$
Taking into account the spin, each subspace contains an electron pair.
Henceforth, we will denote PNOF6($N_{c}$) the method we use, emphasizing
the number $N_{c}$ of usually weakly-occupied orbitals employed in
the description of each electron pair.}

\textcolor{black}{The PNOF6($N_{c}$) energy for a singlet state of
an $N$-electron molecule can be cast as
\begin{equation}
E=\sum\limits _{g=1}^{F}E_{g}+\sum\limits _{f\neq g}^{F}\sum\limits _{p\in\Omega_{f}}\sum\limits _{q\in\Omega_{g}}E_{pq}^{int}\label{PNOF6Nc}
\end{equation}
The first term of the energy (\ref{PNOF6Nc}) draws the system as
independent $F=N/2$ electron pairs described by the following NOF
for two-electron systems,
\begin{equation}
E_{g}=\sum\limits _{p\in\Omega_{g}}n_{p}\left(2\mathcal{H}_{pp}+\mathcal{J}_{pp}\right)+\sum\limits _{p,q\in\Omega_{g},p\neq q}E_{pq}^{int}\label{Eg}
\end{equation}
}

\textcolor{black}{where $\mathcal{H}_{pp}$ is the matrix element
of the kinetic energy and nuclear attraction terms, whereas $\mathcal{J}_{pp}=<pp|pp>$
is the Coulomb interaction between two electrons with opposite spins
at the spatial orbital $p$. It is worth noting that the interaction
energy, the last term of equations (\ref{PNOF6Nc}) and (\ref{Eg}),
is equal for electrons belonging to the same subspace $\Omega_{g}$
or two different subspaces ($\Omega_{g}\neq\Omega_{f}$), therefore,
the intrapair and interpair electron correlations are equally balanced
in PNOF6($N_{c}$). }

\textcolor{black}{The interaction energy $E_{pq}^{int}$ is given
by
\begin{equation}
E_{pq}^{int}=\left(n_{q}n_{p}-\Delta_{qp}\right)\left(2\mathcal{J}_{pq}-\mathcal{K}_{pq}\right)+\Pi_{qp}\mathcal{L}_{pq}\label{Eint}
\end{equation}
}

\textcolor{black}{where $\mathcal{J}_{pq}=\left\langle pq|pq\right\rangle $
and $\mathcal{K}_{pq}=\left\langle pq|qp\right\rangle $ are the usual
direct and exchange integrals, respectively. $\mathcal{L}_{pq}=\left\langle pp|qq\right\rangle $
is the exchange and time-inversion integral \cite{Piris1999}, which
reduces to $\mathcal{K}_{pq}$ for real orbitals. $\Delta$ and $\Pi$
are the auxiliary matrices proposed in reference \cite{Piris2006}
in order to reconstruct the 2-RDM in terms of the occupation numbers.
The conservation of the total spin allows to determine the diagonal
elements as $\Delta_{pp}=n_{p}^{2}$ and $\Pi_{pp}=n_{p}$ \cite{Piris2009},
whereas known analytical necessary $N$-representability conditions
provide bounds for the off-diagonal terms \cite{Piris2010a}. In the
case of PNOF6($N_{c}$), the off-diagonal terms of $\Delta$ and $\Pi$
matrices are
\begin{equation}
\begin{array}{cc|cc|cc}
\Delta_{qp} &  & \Pi_{qp} &  &  & Orbitals\\
\hline e^{-2S}h_{q}h_{p} &  & -e^{-S}\left(h_{q}h_{p}\right)^{\nicefrac{1}{2}} &  &  & q\leq F,p\leq F\\
{\displaystyle \frac{\gamma_{q}\gamma_{p}}{S_{\gamma}}} &  & -\Pi_{qp}^{\gamma} &  &  & \begin{array}{c}
q\leq F,p>F\\
q>F,p\leq F
\end{array}\\
e^{-2S}n_{q}n_{p} &  & e^{-S}\left(n_{q}n_{p}\right)^{\nicefrac{1}{2}} &  &  & q>F,p>F
\end{array}\label{DPi}
\end{equation}
}

\textcolor{black}{where $h_{p}=\left(1-n_{p}\right)$ is the hole
in the spatial orbital $p$. The other magnitudes are defined as
\begin{equation}
\begin{array}{c}
\gamma_{p}=n_{p}h_{p}+\alpha_{p}^{2}-\alpha_{p}S_{\alpha}\\
\\
\alpha_{p}=\begin{cases}
e^{-S}h_{p}\,, & p\leq F\\
e^{-S}n_{p}\,, & p>F
\end{cases}\\
\\
\Pi_{qp}^{\gamma}=\left(n_{q}h_{p}+{\displaystyle \frac{\gamma_{q}\gamma_{p}}{S_{\gamma}}}\right)^{\nicefrac{1}{2}}\left(h_{q}n_{p}+{\displaystyle \frac{\gamma_{q}\gamma_{p}}{S_{\gamma}}}\right)^{\nicefrac{1}{2}}\\
\\
S={\displaystyle \sum_{q=F+1}^{F+FN_{c}}}n_{q},\quad S_{\alpha}={\displaystyle \sum_{q=F+1}^{F+FN_{c}}}\alpha_{q},\quad S_{\gamma}={\displaystyle \sum_{q=F+1}^{F+FN_{c}}}\gamma_{q}
\end{array}
\end{equation}
}

\textcolor{black}{It is noteworthy that the reconstruction of the
2-RDM, and therefore the functional (\ref{PNOF6Nc}), are independent
of the orbital-pairing sum rules (\ref{sumrule_n}). These additional
constraints are imposed to ensure that no fractional electron numbers
appear when non-dynamic electron correlation effects become important
\cite{Piris2011,Matxain2011,Ruiperez2013}. Additionally, this allows
the constraint-free minimization of the PNOF6($N_{c}$) energy (\ref{PNOF6Nc})
with respect to the occupation numbers, which yields substantial savings
of computational time.}

\textcolor{black}{At present, the procedure for the minimization of
the energy (\ref{PNOF6Nc}) with respect to both the occupation numbers
and the natural orbitals is carried out by the iterative diagonalization
method developed by Piris and Ugalde \cite{Piris2009a} implemented
in the DoNOF program package. The matrix element of the kinetic energy
and nuclear attraction terms, as well as the electron repulsion integrals
are inputs to our computational code. In the current implementation,
we have used the GAMESS program \cite{Schmidt1993,Gordon2005} for
this task.}

\subsection{\textcolor{black}{Dipole, Quadrupole and Octupole Moments}}

\textcolor{black}{The potential of the electric field at any point
outside a distribution of charges is simply related to the electric
multipole moments. As any distribution function, the essential features
of the charge distribution can be characterized by its moments, thereby
for an uncharged molecule the first (dipole), second (quadrupole)
and third (octupole) electric moments are the most important terms
in the multipole expansion, therefore, are usually sufficient to characterize
its interaction with an external field. The components of the symmetric
dipole, quadrupole, and octupole moments were defined by Buckingham
\cite{Buckingham1959,Buckingham1967} as}

\textcolor{black}{
\begin{equation}
\text{\ensuremath{{\displaystyle \mu_{\alpha}={\displaystyle -}\int\rho(\mathbf{r}){\displaystyle r_{\alpha}{\displaystyle dV}}}}\ensuremath{{\displaystyle \:+\:\sum_{i=1}^{NUC}Z_{i}R_{i\alpha}}}}\label{eq:dip}
\end{equation}
}

\textcolor{black}{
\begin{equation}
\begin{array}{c}
\text{\ensuremath{{\displaystyle \Theta_{\alpha\beta}=-\frac{1}{2}\int\rho(\mathbf{r})(3{\displaystyle r_{\alpha}{\displaystyle r_{\beta}-\delta_{\alpha\beta}r^{2})dV}}}}}\\
\\
+{\displaystyle \frac{1}{2}\sum_{i=1}^{NUC}Z_{i}(3R_{i\alpha}R_{i\beta}-\delta_{\alpha\beta}R_{i}^{2})}
\end{array}\label{eq:quad}
\end{equation}
}

\textcolor{black}{
\begin{equation}
\begin{array}{c}
{\displaystyle \Omega_{\alpha\beta\gamma}=-\frac{5}{2}\int\rho(\mathbf{r})\, r_{\alpha}r_{\beta}r_{\gamma}dV}\\
\\
+{\displaystyle \frac{1}{2}\:\int\rho(\mathbf{r)}r^{2}\left(r_{\alpha}\delta_{\beta\gamma}+r_{\beta}\delta_{\alpha\gamma}+r_{\gamma}\delta_{\alpha\beta}\right)dV}\\
\\
{\displaystyle +\frac{5}{2}\sum_{i=1}^{NUC}Z_{i}\, R_{i\alpha}R_{i\beta}R_{i\gamma}}\\
\\
{\displaystyle -\frac{1}{2}\sum_{i=1}^{NUC}Z_{i}R_{i}^{2}\left(R_{i\alpha}\delta_{\beta\gamma}+R_{i\beta}\delta_{\alpha\gamma}+R_{i\gamma}\delta_{\alpha\beta}\right)}
\end{array}\label{eq:oct}
\end{equation}
}

\textcolor{black}{where the Greek subscripts denote the Cartesian
directions $\mathit{x,}\, y$ and $\mathit{z}$. Note that the nuclear
contribution is taken into account separately from the electronic
contribution, which arises from the negative charge distribution over
all the space. The formulas (\ref{eq:quad}) and (\ref{eq:oct}) define
symmetric tensors in all subscripts. Moreover, equation (\ref{eq:quad})
defines a traceless tensor for the quadrupole moment, namely, $\Theta_{xx}+\Theta_{yy}+\Theta_{zz}=0$,
similarly, equation (\ref{eq:oct}) leads to $\Omega_{xxz}+\Omega_{yyz}+\Omega_{zzz}=0$
for octupole tensor, and respective permutations between the subscripts
$\mathit{x,}\, y$ and $\mathit{z}$.}

\section{\textcolor{black}{Results and Discussion}}

\textcolor{black}{In the following sections, we show the PNOF6($N_{c}$)
results obtained for the dipole, quadrupole, and octupole moments
with respect to the center of mass. }

\textcolor{black}{The chosen basis set is known to be an important
factor in the calculation of molecular electric properties. We used
the Gaussian basis set of Sadlej \cite{Sadlej1988,Sadlej1991}, which
is a correlation-consistent valence triple-$\zeta$ basis set augmented
with additional basis functions selected specifically for the correlated
calculation of electric properties. Thus, it contains sufficient diffuse
and polarization functions in order to give an accurate description
of the outer-valence region. It has been shown \cite{Hickey2014,Bak_dipoles_CC}
that the Sadlej basis set has effectively the same accuracy as the
aug-cc-pVTZ basis set. }

\textcolor{black}{}
\begin{table*}
\textcolor{black}{\caption{\textcolor{black}{\label{nc_det}$\Theta_{zz}$ component of H$_{2}$
quadrupole moment, in atomic units, obtained by employing PNOF6($N_{c}$)
and CCSD with the Sadlej-pVTZ basis set at the experimental equilibrium
geometry \cite{nist}, together with the experimental value \cite{Orcutt_H2_exp}.}\textcolor{red}{\bigskip{}
}}
}

\noindent \centering{}\textcolor{black}{}%
\begin{tabular}{cccccc}
\hline 
\textcolor{black}{PNOF6($1$)} & \textcolor{black}{PNOF6($3$)} & \textcolor{black}{PNOF6($5$)} & \textcolor{black}{PNOF6($17$)} & \textcolor{black}{CCSD} & \textcolor{black}{EXP.}\tabularnewline
\hline 
\textcolor{black}{$0.3697$} & \textcolor{black}{$0.4030$} & \textcolor{black}{$0.3965$} & \textcolor{black}{$0.3935$} & \textcolor{black}{$0.3935$} & \textcolor{black}{$0.39\pm0.01$}\tabularnewline
\hline 
\end{tabular}
\end{table*}

\textcolor{black}{Since the number $N_{c}$ of usually weakly-occupied
orbitals is related to the description of the electron pairs, we begin
studying the H$_{2}$ molecule, where there are not interpair correlation
effects. This molecule has zero dipole moment, hence the calculated
quadrupole moment values for different $N_{c}$ values are shown in
table \ref{nc_det}.}

\textcolor{black}{As expected, the best description of the electron
pair is obtained when the number of usually weakly-occupied orbitals
is maximum, in fact, the calculated quadrupole moment converges to
the CCSD value, which is the full CI result for this molecule. In
the present work, we will carry out all PNOF6($N_{c}$) calculations
by using the maximum $N_{c}$ value allowed by the Sadlej basis set
for each molecule. In our calculations, we have set to one the occupancies
of the core orbitals. Consequently, the maximum possible value of
$N_{c}$ is given by the number of basis functions above the Fermi
level, divided by the number of the considered strongly occupied orbitals.}

\textcolor{black}{For comparison, we have included the available experimental
data, and the calculated Hartree-Fock (HF) and CCSD values using the
GAMESS program \cite{Schmidt1993,Gordon2005}. The experimental equilibrium
geometries \cite{Bundgen,Junquera,nist,Hirota_ch4geom} have been
used to carry out all calculations. The performance of theoretically
obtained results is established by carrying out a statistical analysis
of the mean absolute errors (MAE) with respect to the experimental
data. Atomic units (a.u.) are used throughout.}

\subsection{\textcolor{black}{Dipole Moment}}

\textcolor{black}{In this work, the dipoles are aligned along the
principal symmetry axis of the studied molecules, set on $z$ direction.
Table \ref{DIP} shows the independent component $\mu_{z}$ of the
dipole moments obtained at the HF, PNOF6($N_{c}$), and CCSD levels
of theory.}

\textcolor{black}{}
\begin{table*}
\textcolor{black}{\caption{$\mu_{z}$ component of molecular dipole moments in atomic units ($ea_{0}$)
computed with the Sadlej-pVTZ basis set at the experimental equilibrium
geometries \cite{nist}. $N_{c}$ is the number of weakly-occupied
orbitals employed in PNOF6$(N_{c})$ for each molecule. \bigskip{}
}
}

\noindent \begin{centering}
\textcolor{black}{}%
\begin{tabular}{>{\raggedright}p{2cm}r>{\raggedleft}p{2cm}c>{\centering}p{2cm}>{\raggedright}p{2cm}}
\hline 
\textcolor{black}{Molecule} & \textcolor{black}{HF$\;\;$} & \textcolor{black}{PNOF6} & \textcolor{black}{$(N_{c})$} & \textcolor{black}{CCSD} & \textcolor{black}{$\;$EXP.}\tabularnewline
\hline 
\textcolor{black}{HF} & \textcolor{black}{$0.7565$} & \textcolor{black}{$0.7223$} & \textcolor{black}{$7$} & \textcolor{black}{$0.6994$} & \textcolor{black}{$0.7089$ \cite{Bundgen}}\tabularnewline
\textcolor{black}{BH$^{*}$} & \textcolor{black}{$0.6854$} & \textcolor{black}{$0.5395$} & \textcolor{black}{$38$} & \textcolor{black}{$0.5551$} & \textcolor{black}{$0.4997$ \cite{Bundgen}}\tabularnewline
\textcolor{black}{H$_{2}$O} & \textcolor{black}{$0.7808$} & \textcolor{black}{$0.7458$} & \textcolor{black}{$9$} & \textcolor{black}{$0.7225$} & \textcolor{black}{$0.7268$ \cite{Bundgen}}\tabularnewline
\textcolor{black}{H$_{2}$CO} & \textcolor{black}{$1.1134$} & \textcolor{black}{$0.9872$} & \textcolor{black}{$10$} & \textcolor{black}{$0.9084$} & \textcolor{black}{$0.9175$ \cite{Bundgen}}\tabularnewline
\textcolor{black}{HCl} & \textcolor{black}{$0.4746$} & \textcolor{black}{$0.4598$} & \textcolor{black}{$8$} & \textcolor{black}{$0.4416$} & \textcolor{black}{$0.4301$ \cite{Bundgen}}\tabularnewline
\textcolor{black}{HCCF} & \textcolor{black}{$0.3535$} & \textcolor{black}{$0.3189$} & \textcolor{black}{$9$} & \textcolor{black}{$0.2733$} & \textcolor{black}{$0.2872$ \cite{FlygareEXP}}\tabularnewline
\textcolor{black}{NH$_{3}$} & \textcolor{black}{$0.6372$} & \textcolor{black}{$0.6153$} & \textcolor{black}{$12$} & \textcolor{black}{$0.5943$} & \textcolor{black}{$0.5789$ \cite{Junquera}}\tabularnewline
\textcolor{black}{PH$_{3}$} & \textcolor{black}{$0.2780$} & \textcolor{black}{$0.2755$} & \textcolor{black}{$13$} & \textcolor{black}{$0.2340$} & \textcolor{black}{$0.2258$ \cite{Siegfried}}\tabularnewline
\textcolor{black}{O$_{3}$} & \textcolor{black}{$0.3033$} & \textcolor{black}{$0.1370$} & \textcolor{black}{$7$} & \textcolor{black}{$0.2276$} & \textcolor{black}{$0.2099$ \cite{Maroulis_ozone}}\tabularnewline
\textcolor{black}{ClF} & \textcolor{black}{$0.4453$} & \textcolor{black}{$0.3226$} & \textcolor{black}{$6$} & \textcolor{black}{$0.3451$} & \textcolor{black}{$0.3462$ \cite{Fabricant_ClF}}\tabularnewline
\textcolor{black}{CH$_{3}$F} & \textcolor{black}{$0.7706$} & \textcolor{black}{$0.7283$} & \textcolor{black}{$10$} & \textcolor{black}{$0.6919$} & \textcolor{black}{$0.7312$ \cite{Russel_CH3F}}\tabularnewline
\textcolor{black}{CH$_{3}$CCH} & \textcolor{black}{$0.3203$} & \textcolor{black}{$0.3141$} & \textcolor{black}{$12$} & \textcolor{black}{$0.2866$} & \textcolor{black}{$0.3070$ \cite{Arapiraca_ch3cch}}\tabularnewline
\textcolor{black}{CO} & \textcolor{black}{$-0.0987$} & \textcolor{black}{$0.0414$} & \textcolor{black}{$9$} & \textcolor{black}{$0.0725$} & \textcolor{black}{$0.0481$ \cite{Bundgen}}\tabularnewline
\hline 
\textcolor{black}{MAE} & \textcolor{black}{$0.0843$} & \textcolor{black}{$0.0309$} &  & \textcolor{black}{$0.0177$} & \tabularnewline
\end{tabular}\textcolor{black}{\label{DIP}\bigskip{}
}
\par\end{centering}

\noindent \raggedright{}\textcolor{black}{$^{*}$Calculations performed
with the aug-cc-pVTZ basis set.}
\end{table*}

\textcolor{black}{Overall, the inclusion of electron correlation effects
through, both PNOF6($N_{c}$) and CCSD, improves significantly the
performance of the HF method. PNOF6($N_{c}$) and CCSD afford MAEs
with respect to experimental data of 0.0309 a.u. and 0.0177 a.u.,
respectively. It is worth noting the agreement between PNOF6($N_{c}$)
and CCSD results, as well as with the experimental data. Note that
the aug-cc-pVTZ basis set of Dunning \cite{Dunning1989} was used
for the BH molecule since there is no Sadlej-pVTZ basis set available
for Boron. In this case, the PNOF6($38$) result is very close to
the Full-CI/aug-cc-pVTZ value obtained by Halkier et al. \cite{HalkierBH_HF},
0.5433 a.u., showing a result as good as the CCSD one.}

\textcolor{black}{The electronic structure and bonding situation of
carbon monoxide is of special interest for modern electronic structure
methods. The dipole moment of CO, extensively studied in Refs. \cite{Frenking2006,Bak_dipoles_CC,Maroulis_CO},
is very small (0.0481 a.u.) and ends at the carbon atom, although
carbon is less electronegative than oxygen. The result shown in Table
\ref{DIP} is representative, while HF gives the wrong direction for
the CO dipole moment, PNOF6($9$) corrects the sign, giving a result
that is in excellent agreement with the experimental value. Remarkably,
the result obtained at CCSD level is $34\%$ away from the experimental
value, so that it is necessary to include third order triplet excitations
in the cluster theory in order to obtain a reasonable value, such
as the one reported by Maroulis \cite{Maroulis_CO} at the CCSD(T)
level, 0.0492 a.u.. Accordingly, the relevant electron correlation
for CO is well accounted by the PNOF6($9$) method.}

\textcolor{black}{Regarding the values obtained for HF, H$_{2}$O,
H$_{2}$CO, HCl, NH$_{3}$, and ClF, PNOF6($N_{c}$) competes with
Coupled Cluster, providing values that differ from experimental data
in less than a $7\%$. In the case of HCCF and PH$_{3}$, PNOF6($N_{c}$)
seems to lack relevant dynamic electron correlation and thereby the
obtained dipole moments are not as accurate as the CCSD ones. Conversely,
our values are in excellent agreement with experimental data in the
case of the methyls CH$_{3}$F and CH$_{3}$CCH, often attached to
large organic molecules, giving dipole moments with errors of $0.4\%$
and $2\%$ respectively, with respect to experimental values.}

\textcolor{black}{A special case is ozone, which is a molecule with
strong multiconfigurational character. The PNOF6(7) dipole moves into
the right direction from the HF value, but overestimates the effects
of the electron correlation. Taking into account the good CCSD result
for O$_{3}$, which is not valid for higher electric moments, it seems
that the dynamic electron correlation compensates for the lack of
non-dynamical in this method, and could improve our numerical value
of the dipole.}

\subsection{\textcolor{black}{Quadrupole Moment}}

\textcolor{black}{Tables \ref{quad-1} and \ref{quad-2} list the
molecular quadrupole moments obtained at the HF, CCSD, MRSD-CI and
PNOF6($N_{c}$) levels of theory, along with the experimental values
taken from Refs. \cite{Bundgen,Junquera,Fabricant_ClF,FlygareEXP,Maroulis_ozone,Russel_CH3F,Siegfried,heard_benzene,luca_c2h6,Orcutt_H2_exp,Dorothy}.
Inspection of these Tables shows that PNOF6($N_{c}$) quadrupole moments
agree satisfactorily with the experimental data, whereas the discrepancies
are consistent with those observed using the CCSD and MRSD-CI methods
in most cases.}

\textcolor{black}{}
\begin{table*}
\textcolor{black}{\caption{$\Theta_{zz}$ component of the quadrupole moments, in atomic units,
computed with the Sadlej-pVTZ basis set at the experimental equilibrium
geometries \cite{nist} for molecules with linear, $C_{3v}$, $D_{6h}$,
and $D_{3d}$ symmetry. $N_{c}$ is the number of weakly-occupied
orbitals employed in PNOF6$(N_{c})$ for each molecule.\bigskip{}
}
}

\noindent \begin{centering}
\textcolor{black}{}%
\begin{tabular}{>{\raggedright}p{2cm}c>{\raggedleft}p{2cm}c>{\centering}p{2cm}>{\raggedright}p{3cm}}
\hline 
\textcolor{black}{Molecule} & \textcolor{black}{HF} & \textcolor{black}{PNOF6} & \textcolor{black}{($N_{c}$)} & \textcolor{black}{CCSD} & \textcolor{black}{$\quad\;$EXP.}\tabularnewline
\hline 
\textcolor{black}{H$_{2}$} & \textcolor{black}{$0.4381$} & \textcolor{black}{$0.3935$} & \textcolor{black}{$17$} & \textcolor{black}{$0.3935$} & \textcolor{black}{$0.39\pm0.01$ \cite{Orcutt_H2_exp}}\tabularnewline
\textcolor{black}{HF} & \textcolor{black}{$1.7422$} & \textcolor{black}{$1.6939$} & \textcolor{black}{$7$} & \textcolor{black}{$1.7156$} & \textcolor{black}{$1.75\pm0.02$ \cite{Junquera}}\tabularnewline
\textcolor{black}{BH$^{*}$} & \textcolor{black}{$2.6772$} & \textcolor{black}{$2.3706$} & \textcolor{black}{$38$} & \textcolor{black}{$2.3388$} & \textcolor{black}{$\quad2.3293^{\dagger}$ \cite{HalkierBH_HF}}\tabularnewline
\textcolor{black}{HCl} & \textcolor{black}{$2.8572$} & \textcolor{black}{$2.7753$} & \textcolor{black}{$8$} & \textcolor{black}{$2.7233$} & \textcolor{black}{$2.78\pm0.09$ \cite{Bundgen}}\tabularnewline
\textcolor{black}{HCCF} & \textcolor{black}{$3.3530$} & \textcolor{black}{$3.2482$} & \textcolor{black}{$9$} & \textcolor{black}{$2.9335$} & \textcolor{black}{$2.94\pm0.10$ \cite{FlygareEXP}}\tabularnewline
\textcolor{black}{CO} & \textcolor{black}{$1.5366$} & \textcolor{black}{$1.4562$} & \textcolor{black}{$9$} & \textcolor{black}{$1.4889$} & \textcolor{black}{$1.44\pm0.30$ \cite{Junquera}}\tabularnewline
\textcolor{black}{N$_{2}$} & \textcolor{black}{$0.9397$} & \textcolor{black}{$1.0530$} & \textcolor{black}{$9$} & \textcolor{black}{$1.1712$} & \textcolor{black}{$1.09\pm0.07$ \cite{Junquera}}\tabularnewline
\textcolor{black}{NH$_{3}$} & \textcolor{black}{$2.1258$} & \textcolor{black}{$2.1080$} & \textcolor{black}{$12$} & \textcolor{black}{$2.1661$} & \textcolor{black}{$2.45\pm0.30$ \cite{Bundgen}}\tabularnewline
\textcolor{black}{PH$_{3}$} & \textcolor{black}{$1.7217$} & \textcolor{black}{$1.6507$} & \textcolor{black}{$13$} & \textcolor{black}{$1.5695$} & \textcolor{black}{$1.56\pm0.70$ \cite{Siegfried}}\tabularnewline
\textcolor{black}{ClF} & \textcolor{black}{$0.9413$} & \textcolor{black}{$1.1122$} & \textcolor{black}{$6$} & \textcolor{black}{$1.0514$} & \textcolor{black}{$1.14\pm0.05$ \cite{Fabricant_ClF}}\tabularnewline
\textcolor{black}{CH$_{3}$F} & \textcolor{black}{$0.3482$} & \textcolor{black}{$0.3269$} & \textcolor{black}{$10$} & \textcolor{black}{$0.3002$} & \textcolor{black}{$0.30\pm0.02$ \cite{Russel_CH3F}}\tabularnewline
\textcolor{black}{C$_{2}$H$_{2}$} & \textcolor{black}{$5.3655$} & \textcolor{black}{$5.1531$} & \textcolor{black}{$12$} & \textcolor{black}{$4.6850$} & \textcolor{black}{$4.71\pm0.14$ \cite{Dorothy}}\tabularnewline
\textcolor{black}{C$_{2}$H$_{6}$} & \textcolor{black}{$0.6329$} & \textcolor{black}{$0.6275$} & \textcolor{black}{$13$} & \textcolor{black}{$0.6234$} & \textcolor{black}{$0.59\pm0.07$ \cite{luca_c2h6}}\tabularnewline
\textcolor{black}{C$_{6}$H$_{6}$} & \textcolor{black}{$6.6418$} & \textcolor{black}{$6.3571$} & \textcolor{black}{$12$} & \textcolor{black}{$5.6653$} & \textcolor{black}{$6.30\pm0.27$ \cite{heard_benzene}}\tabularnewline
\textcolor{black}{CH$_{3}$CCH} & \textcolor{black}{$4.2913$} & \textcolor{black}{$4.1146$} & \textcolor{black}{$12$} & \textcolor{black}{$3.6939$} & \textcolor{black}{$3.58\pm0.01$ \cite{nist}}\tabularnewline
\textcolor{black}{CO$_{2}$} & \textcolor{black}{$3.8087$} & \textcolor{black}{$3.6012$} & \textcolor{black}{$8$} & \textcolor{black}{$3.1966$} & \textcolor{black}{$3.19\pm0.13$ \cite{Junquera}}\tabularnewline
\hline 
\textcolor{black}{MAE} & \textcolor{black}{$0.2646$} & \textcolor{black}{$0.1517$} &  & \textcolor{black}{$0.0902$} & \tabularnewline
\end{tabular}\textcolor{black}{\label{quad-1}}
\par\end{centering}

\noindent \begin{raggedright}
\textcolor{black}{$^{*}$Calculations performed with the aug-cc-pVTZ
basis set.}
\par\end{raggedright}

\noindent \raggedright{}\textcolor{black}{$^{\dagger}$ Full CI calculation
reported by Halkier et al. \cite{HalkierBH_HF}}
\end{table*}

\textcolor{black}{}
\begin{table*}
\textcolor{black}{\caption{$\Theta_{zz}$ and $\Theta_{xx}$ components of molecular quadrupole
moments, in atomic units, computed using the Sadlej-pVTZ basis set
at the experimental equilibrium geometries \cite{nist}. $N_{c}$
is the number of weakly-occupied orbitals employed in PNOF6$(N_{c})$
for each molecule.\bigskip{}
}
}

\noindent \centering{}\textcolor{black}{}%
\begin{tabular}{>{\raggedright}p{2.5cm}r>{\raggedleft}p{2cm}c>{\centering}p{3cm}>{\raggedright}p{3cm}}
\hline 
\textcolor{black}{Molecule} & \textcolor{black}{HF$\;\;$} & \textcolor{black}{PNOF6} & \textcolor{black}{($N_{c}$)} & \textcolor{black}{MRSD-CI} & \textcolor{black}{$\quad\;$EXP.}\tabularnewline
\hline 
\textcolor{black}{H$_{2}$O ($xx$)} & \textcolor{black}{$1.7966$} & \textcolor{black}{$1.7808$} & \textcolor{black}{$9$} & \textcolor{black}{$1.805$0} & \textcolor{black}{$1.86\pm0.02$ \cite{Bundgen}}\tabularnewline
\textcolor{black}{H$_{2}$O ($zz$)} & \textcolor{black}{$0.0981$} & \textcolor{black}{$0.0869$} & \textcolor{black}{$9$} & \textcolor{black}{$0.0950$} & \textcolor{black}{$0.10\pm0.02$ \cite{Bundgen}}\tabularnewline
\textcolor{black}{H$_{2}$CO ($xx$)} & \textcolor{black}{$0.1019$} & \textcolor{black}{$0.0516$} & \textcolor{black}{$10$} & \textcolor{black}{$0.1100$} & \textcolor{black}{$0.04\pm0.12$ \cite{kukolich_h2co}}\tabularnewline
\textcolor{black}{H$_{2}$CO ($zz$)} & \textcolor{black}{$0.0921$} & \textcolor{black}{$0.1255$} & \textcolor{black}{$10$} & \textcolor{black}{$0.2230$} & \textcolor{black}{$0.20\pm0.15$ \cite{kukolich_h2co}}\tabularnewline
\textcolor{black}{C$_{2}$H$_{4}$ ($xx$)} & \textcolor{black}{$2.7819$} & \textcolor{black}{$2.5892$} & \textcolor{black}{$13$} & \textcolor{black}{$2.3700$} & \textcolor{black}{$2.45\pm0.12$ \cite{Bundgen}}\tabularnewline
\textcolor{black}{C$_{2}$H$_{4}$ ($zz$)} & \textcolor{black}{$1.4942$} & \textcolor{black}{$1.3266$} & \textcolor{black}{$13$} & \textcolor{black}{$1.1700$} & \textcolor{black}{$1.49\pm0.11$ \cite{Bundgen}}\tabularnewline
\textcolor{black}{O$_{3}$ ($xx$)} & \textcolor{black}{$1.1175$} & \textcolor{black}{$1.2426$} & \textcolor{black}{$7$} & \textcolor{black}{$1.2830$} & \textcolor{black}{$1.03\pm0.12$ \cite{Maroulis_ozone}}\tabularnewline
\textcolor{black}{O$_{3}$ ($zz$)} & \textcolor{black}{$-0.2387$} & \textcolor{black}{$0.3606$} & \textcolor{black}{$7$} & \textcolor{black}{$0.1680$} & \textcolor{black}{$0.52\pm0.08$ \cite{Maroulis_ozone}}\tabularnewline
\hline 
\textcolor{black}{MAE} & \textcolor{black}{$0.1772$} & \textcolor{black}{$0.1066$} &  & \textcolor{black}{$0.1448$} & \tabularnewline
\end{tabular}\textcolor{black}{\label{quad-2}}
\end{table*}

\textcolor{black}{In the case of linear molecules (H$_{2}$, HF, BH,
HCl, HCCF, ClF, CO, C$_{2}$H$_{2}$, CO$_{2}$ and N$_{2}$), NH$_{3}$
and PH$_{3}$, belonging to the $C_{3v}$ point symmetry group, the
$D_{6h}$ C$_{6}$H$_{6}$ molecule, and the trigonal planar C$_{2}$H$_{6}$,
which has $D_{3d}$ symmetry, the relation $\Theta_{xx}=\Theta_{yy}=-\frac{1}{2}\Theta_{zz}$
holds for quadrupole moment tensor, so $\Theta_{zz}$ alone is sufficient
to determine completely the quadrupole moment. Setting the main axis
of symmetry in the z direction of the coordinate system, the results
for these molecules are reported in Table \ref{quad-1}. From the
latter, one can observe that PNOF6($N_{c}$) yields a MAE of 0.15
a.u., hence considering the added complexity of the quadrupole moment,
the performance of PNOF6($N_{c}$) is within a reasonable accuracy. }

\textcolor{black}{Taking into account the experimental uncertainty,
PNOF6($N_{c}$) results agree with the experimental data for H$_{2}$,
HCl, CO, N$_{2}$, PH$_{3}$, ClF, CH$_{3}$F, C$_{2}$H$_{6}$, and
C$_{6}$H$_{6}$. The value obtained for H$_{2}$ reproduces the experimental
one with high precision. It is also worth noting the excellent agreement
with the experiment obtained for the quadrupole moment of Benzene,
which is of great interest for many fields of chemistry and biology
\cite{heard_benzene,shimizu_benzene}. Indeed, the quadrupole moment
of Benzene plays an important role in determining the crystal structures
and molecular recognition in biological systems because it is the
key to the intermolecular interactions between $\pi$-systems.}

\textcolor{black}{For HCCF, NH$_{3}$, C$_{2}$H$_{2}$, and CO$_{2}$,
the quadrupole moments fall out of the experimental error intervals,
however, in the case of HCCF, C$_{2}$H$_{2}$, and CO$_{2}$ the
mean relative percentage error is below $11\%$, whereas the results
obtained for NH$_{3}$ is only 0.05 a.u. away from the higher limit
of the experimental uncertainty. For CH$_{3}$CCH the PNOF6($12$)
result deviates from the experimental value in a $13\%$, ergo more
dynamic correlation is clearly necessary to improve this result, an
effect not observed for the dipole moment of this molecule.}

\textcolor{black}{For the Hydrogen fluoride, the HF result is the
closest to the experimental value, however, the PNOF6($7$) result
is in outstanding agreement with the full-CI/aug-cc-pVTZ value of
1.6958 a.u. \cite{HalkierBH_HF}. For the Boron monohydride, the experimental
quadrupole moment is not available, so we use the full-CI/aug-cc-pVTZ
calculation reported by Halkier et al. \cite{HalkierBH_HF}, 2.3293
a.u., in order to carry out the comparison. The agreement between
PNOF6(38) and full-CI is good, according to the relative percentage
error obtained below 1.7\%.}

\textcolor{black}{Table \ref{quad-2} shows the $\Theta_{zz}$ and
$\Theta_{xx}$ components obtained for H$_{2}$O, H$_{2}$CO, C$_{2}$H$_{4}$,
and O$_{3}$. In this work, we use the traceless quadrupole moment,
hence two components are sufficient to determine completely this magnitude.
On the other hand, MRSD-CI values are significantly better than CCSD
calculations when many components of the quadrupole tensor are studied
\cite{Bundgen}, thereby MRSD-CI is used as benchmark theoretical
method in Table \ref{quad-2}.}

\textcolor{black}{According to the results reported in Table \ref{quad-2},
PNOF6($N_{c}$) performs better than the MRSD-CI method for this selected
set of molecules. For H$_{2}$O and H$_{2}$CO, the PNOF6($N_{c}$)
values fall into the experimental error interval, which is specially
broad for H$_{2}$CO. In the case of the C$_{2}$H$_{4}$ molecule,
the longitudinal component $\Theta_{zz}$ obtained with PNOF6(13)
is near the limit of the experimental error interval, as well as the
$\Theta_{xx}$ component. Finally, we have the results obtained for
O$_{3}$, which is a stringent test for quadrupole calculations due
to its two-configurational character \cite{Maroulis_ozone,Watts_ozone}.
One can observe that Ozone is well described by PNOF6($7$) comparing
to the results obtained by using HF and MRSD-CI methods.}

\subsection{\textcolor{black}{Octupole Moment}}

\textcolor{black}{The octupole moment is particularly interesting
in the case of methane. The octupole moment is the first non-zero
term in the multipole expansion of the electrostatic interaction for
methane molecule, so it is crucial in order to describe properly its
interactions with external fields. Actually, the octupole-octupole
interaction is the main long-range orientation dependent interaction
in methane. Moreover, the complex charge distribution of methane,
which has long been studied in the literature \cite{Coe_CH4,hendrik_methane,Hirota_ch4geom},
is mainly dependent on its octupole moment, thus, the octupole moment
is essential to characterize the charge distribution of tetrahedral
molecules.}

\textcolor{black}{For tetrahedral molecules the octupole moment is
simply given by one component, namely $\Omega=\Omega_{xyz}$. Employing
PNOF6(14) with the Sadlej-pVTZ basis set at the experimental equilibrium
geometry \cite{Hirota_ch4geom}, the result obtained for CH$_{4}$
is $\Omega_{xyz}=2.1142\: a.u.$, whereas the experimental mark reported
in Ref. \cite{CohenEXP} is $\Omega_{xyz}=2.95\pm0.17\: a.u.$ Although
the PNOF6(14) result falls out of the experimental interval error,
this value is reasonable taking into account the discrepancies between
experimental marks obtained by different experimental techniques \cite{CohenEXP}.
Besides, comparing to theoretical calculations, the PNOF6(14) value
is very close to the result obtained by using CCSD, $\Omega_{xyz}=2.0595\: a.u.$.
Consequently, we can conclude that PNOF6(14) describes properly the
octupole moment of methane.}

\section{\textcolor{black}{Conclusions}}

\textcolor{black}{The PNOF6 method, in its extended version, has been
assessed by comparing the molecular electric moments with the experimental
data as well as with CCSD and MRSD-CI theoretical values. The dipole,
quadrupole and octupole moments for a selected set of well-characterized
21 molecules have been calculated at the experimental equilibrium
geometries using the triple-$\zeta$ Gaussian basis set with polarization
functions developed by Sadlej. Our results show that PNOF6($N_{c}$)
is able to predict electric properties as accurate as high-level electronic
structure methods such as CCSD or MRSD-CI, therefore the functional
computes quite accurately the charge distribution of molecular systems.
To our knowledge, this is the first NOF study of higher multipole
moments such as quadrupole and octupole moments.}

\textcolor{black}{For PNOF6($N_{c}$) dipole moments, the obtained
MAE with respect to experimental data is $0.0309\: a.u.$, being consistent
with the theoretical benchmark calculations. Remarkable is the result
obtained by PNOF6($9$) for Carbon monoxide, for which, HF gives a
wrong direction of the dipole and CCSD overestimates it severely,
whereas PNOF6($9$) corrects the sign, giving a result that is in
excellent agreement with the experimental mark.}

\textcolor{black}{The high performance of PNOF6($N_{c}$) in computing
electric quadrupole moments has been shown by most of the studied
molecules, for which the computed values fall into the experimental
interval error. It has been shown that the method is capable of providing
the different components of the quadrupole moment tensor. The PNOF6($N_{c}$)
MAE with respect to the experiment is $0.1291\: a.u.$, which is very
close to the corresponding MAEs of $0.0902\: a.u.$ and $0.1448\: a.u.$
obtained by using the well-established CCSD and MRSD-CI methods, respectively.
In particular, the results obtained for the ozone molecule with a
marked multiconfigurational character, show that the method is able
to treat properly non-dynamic and dynamic electron correlations.}

\textcolor{black}{Finally, the study of the octupole moment was focused
here on methane, due to its important role in the description of the
long-range electrostatic interactions for this molecule. The PNOF6(14)
result is in excellent agreement with the value provided by the CCSD
method.}

\selectlanguage{american}%

\section*{\textcolor{black}{Acknowledgements}}

\textcolor{black}{Financial support comes from Eusko Jaurlaritza (Ref.
IT588-13) and Ministerio de Economía y Competitividad (Ref. CTQ2015-67608-P).
The authors thank for technical and human support provided by IZO-SGI
SGIker of UPV/EHU and European funding (ERDF and ESF). One of us (I.M.)
is grateful to Vice-Rectory for research of the UPV/EHU for the PhD.
grant, and also to Dr. Mauricio Rodríguez for helpful discussions.}

\selectlanguage{english}%
\textcolor{black}{\smallskip{}
}

\textcolor{black}{\bibliographystyle{apsrev}
\addcontentsline{toc}{section}{\refname}\bibliography{manuscript}
\newpage{}}

\includegraphics[scale=0.7]{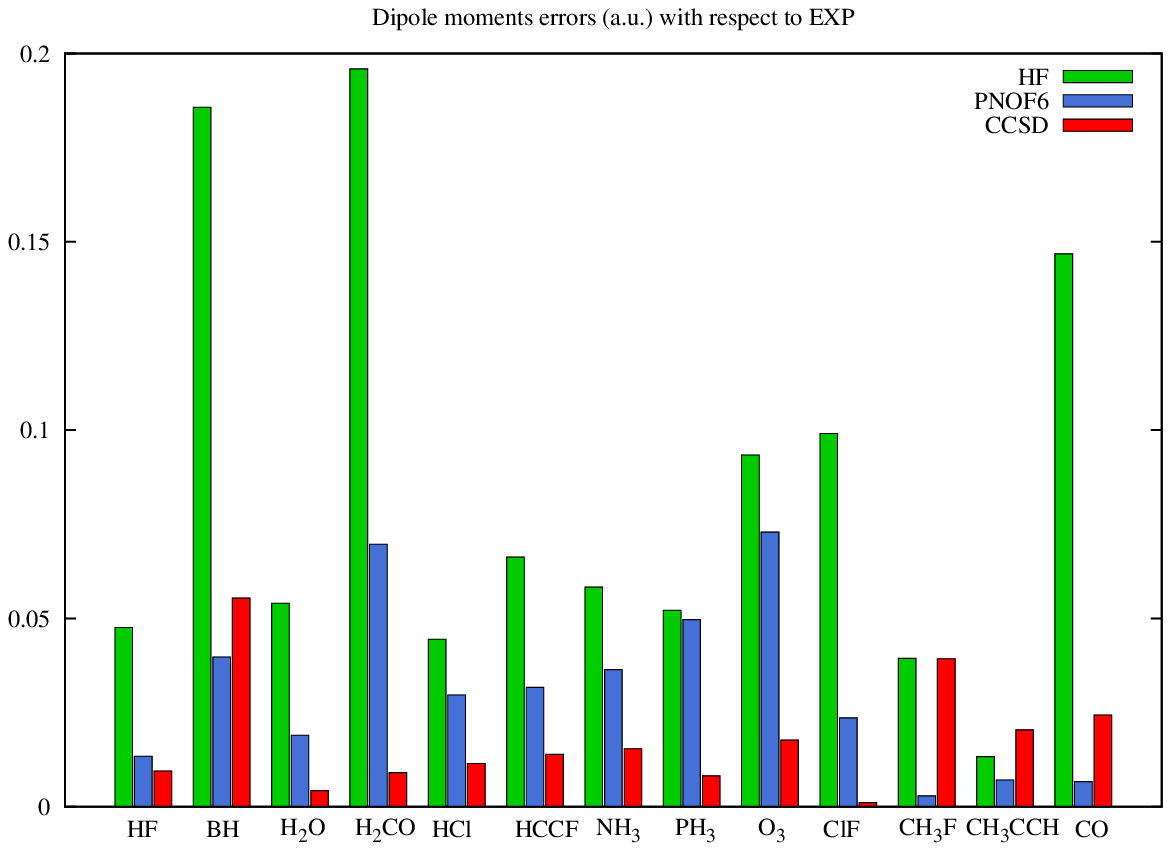}

\includegraphics[scale=0.7]{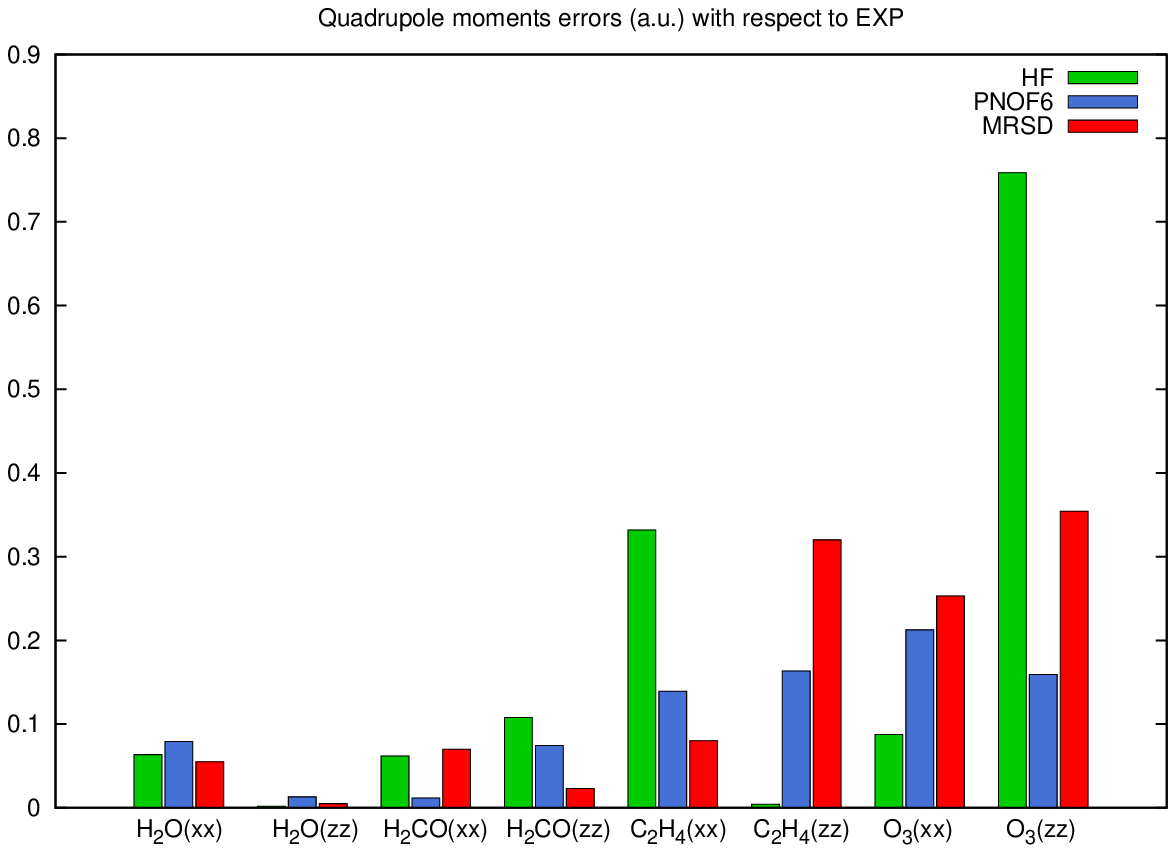}

\includegraphics[scale=0.7]{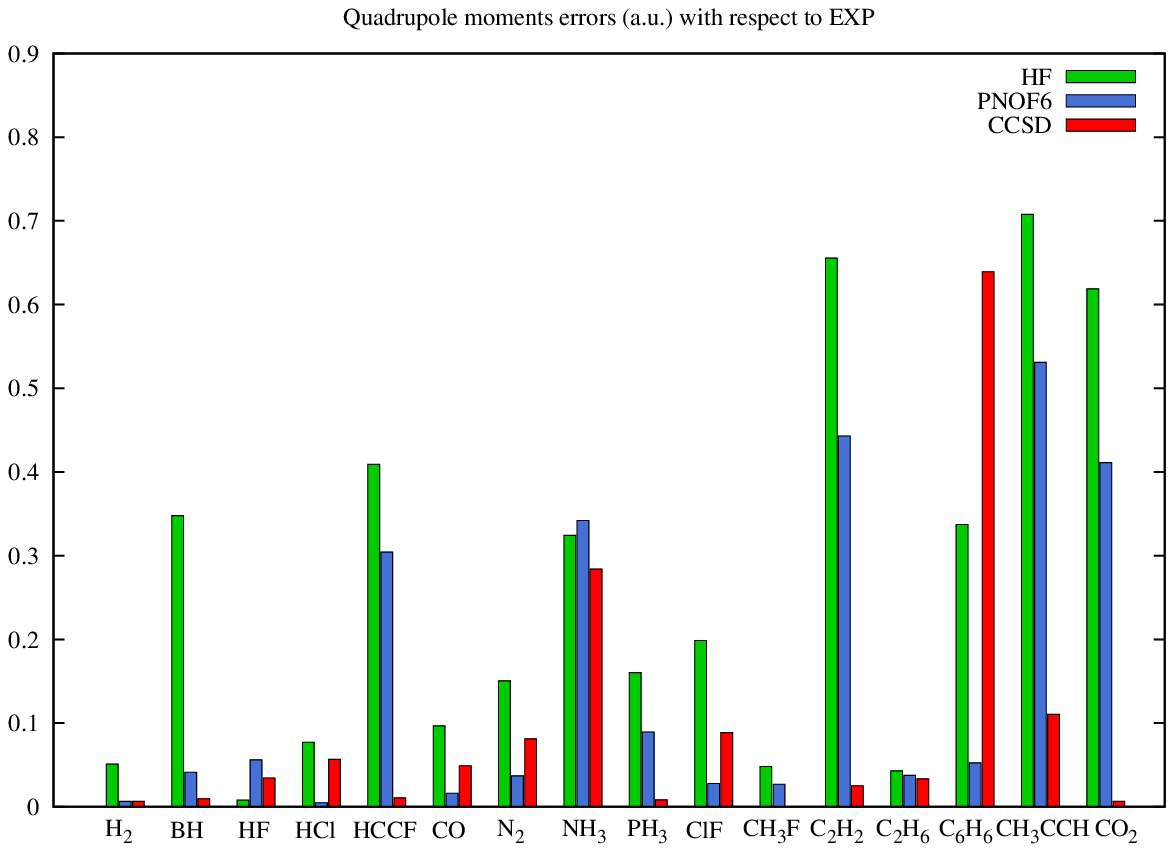}

\end{document}